\documentclass{PoS}

\title{Quasi-spherical accretion in X-ray pulsars}

\ShortTitle{Quasi-spherical accretion}

\author{\speaker{K.~Postnov}$^a$, N.~Shakura$^a$,  A.~Kochetkova$^a$, L.
Hjalmarsdotter$^a$ \\
    \llap{$^a$}Sternberg Astronomical Institute, 13, Universitetskij pr., 119992 Moscow, Russia\\
E-mail: \email{kpostnov@gmail.com}, \email{nikolai.shakura@gmail.com},
\email{apostnova@mail.com}, 
\email{astrogirl@telia.com}}

\abstract{
Quasi-spherical accretion in wind-fed X-ray pulsars is discussed. 
At X-ray luminosities $<4\times 10^{36}$~erg/s, a hot convective shell is 
formed around the neutron star magnetosphere, and subsonic
settling accretion regime sets in. In this regime, accretion rate
onto neutron star is determined by the ability of plasma
to enter magnetosphere via Rayleigh-Taylor instability. 
A gas-dynamic theory of settling accretion
is constructed taking into account
anisotropic turbulence. The angular momentum can be
transferred through the quasi-static shell via
large-scale convective motions initiating turbulence cascade. 
The angular velocity distribution in the shell is found
depending on the turbulent viscosity prescription.
Comparison with observations of long-period X-ray
wind-fed pulsars shows that  an almost
iso-angular-momentum distribution is most likely realized in their shells. 
The theory explains long-term spin-down in wind-
fed accreting pulsars (e.g. GX 1+4) and properties of short-term
torque-luminosity correlations. The theory can be applied to 
slowly rotating low-luminosity X-ray pulsars and non-stationary 
accretion phenomena observed in some SFXTs. 
}

\FullConference{A scientific workshop in Sardinia ''The extreme and variable high X-ray
sky'',\\
September 19-23, 2011\\
Chia Laguna, Sardegna, Italy, }

\begin{document}

\section{Introduction}
Accreting magnetized neutron stars (NS) are observed as X-ray pulsars
(XPSR). The observed characteristics of XPSRs are their X-ray luminosity $L_x$, pulse
period $P^*$ (or angular frequency $\Omega^*=2\pi/P^*$) and its first time
derivative $\dot P^*$ (or $\dot \omega^*$), the pulse profile at different
energy bands, and X-ray spectrum. The timing properties (i.e., the temporal
behavior of $\dot\omega^*$) and their correlation with variations of $L_x$
provide important clues on the interaction mechanism of accreting plasma
with the rotating NS magnetosphere \cite{Bildsten_ea97}. Spectral features
(especially, cyclotron lines) allow direct probing the strength 
of the near-surface magnetic field of NSs \cite{Staubert03}.  Their
dependence on variable X-ray luminosity in persistent (Her X-1,
\cite{Staubert_ea07}) and transient (e.g. 4U0115+63, \cite{Tsygankov_ea07}) 
XPSRs provides information on the accretion column properties
(its geometry etc.).

XPSRs can be persistent (Her X-1, Cen X-3,
Vela X-1, millisecond XPSRs, etc.) or transient (A0535+26, 4U0115+63, etc.), and
are      
found in both high-mass and low-mass X-ray binaries (HMXB: Cen X-3, Vela
X-1, transients, LMXB: Her X-1, 4U1626-67, millisecond XPSRs). 
The matter
from the secondary star can be supplied to NS via Roche lobe
overflow (the most frequent case in LMXBs) or can be captured from stellar
wind of the optical companion (in HMXB). In the first
case, the specific angular momentum of accreting matter $j_m$ is  usually
high enought for an accretion disk to be formed around the NS magnetosphere,
i.e. $j_m>j_K(R_A)=\sqrt{GMR_A}$, where $R_A$ is the characteristic radius
of magnetopshere (the Alfven radius), $M$ is the NS mass, $G$ is the Newton
gravity constant\footnote{In the extreme case of very large magnetospheres,
$R_A\sim a$, where $a$ is the orbital radius, the flow does not form a disk;
this case is realized in AM Her binaries (polars).}. 

\section{Two regimes of quasi-spherical accretion}

In the case of accretion
from the wind considered here, the accretion flow geometry can be more complicated: the
condition for the disk formation  $j_m>j_K(R_A)$ can be satisfied or not,
again depending on the size of the magnetosphere $R_A$ and on the specific
angular momentum of gravitationally captured matter from the stellar wind.
The last quantity can be expressed through the characteristic gravitational
capture radius $R_B=2GM/v^2$(the Bondi radius), where $v^2=v_w^2+v_{orb}^2$ is the relative
velocity of wind (moving with velocity $v_w$)and the NS (moving with orbital
velocity $v_{orb}$ through the wind) as $j_m=k \Omega_b R_B^2$. Here
$\Omega_b$ is the orbital angular frequency, $k$ is the numerical
coefficient which can be positive (prograde $j_m$)
or negative (retrograde $j_m$) \cite{Ho_ea1989}, depending on properties 
of (generally inhomgeneous) stellar wind. 

There can be two
different regimes of quasi-spherical accretion. The captured matter heated up
in the bow shock  at $\sim R_B$ to high temperatures $k_BT\sim m_p v^2$. If
the characteristic cooling time of plasma $t_{cool}$ is smaller than the
free-fall time $t_{ff}= R_B/\sqrt{2GM/R_B}$, the gas falls
supersonically toward the magnetosphere with the formation of a shock. 
This regime is usually considered in connection with bright XPSRs 
\cite{AronsLea76}, \cite{Burnard_ea83}. The role of X-ray photons
generated near the NS surface is two-fold: first, they can heat up plasma in the
bow-shock zone via photoionization, and second, 
they cool down the
hot plasma near the magnetosphere (with $k_BT\sim GM/R_A$) by Compton
processes thus allowing matter to enter the
magnetosphere via the Rayleigh-Taylor instability \cite{ElsnerLamb77}.

In the free-fall accretion regime, the X-ray luminosity (the mass
accretion rate $\dot M$) is determined by the rate of gravitational capture
of stellar wind at $R_B$. The accretion torque applied to NS due to plasma-magnetopshere
interaction is always of the same sign as the specific angular momentum of
the gravitationally captured stellar wind, i.e. the NS can spin-up or spin
down depending on the prograde or retrograde $j_m$. 

If 
the relative wind velocity $v$ at $R_B$ is slow ($\lesssim 80$~km/s), the  
photoionization heating of plasma is important, but the radiation cooling time of plasma is
shorter than the free-fall time, so the free-fall accretion regime is realized.
If the wind velocity is larger than $\sim 80-100$~km/s, the post-shock
temperature is higher than $5 \times 10^5$~K (the maximum photoionization
temperature for a photon temperature of several keV); for $L_x\lesssim 4\times 10^{36}$~erg/s,
the plasma radiative cooling time is longer than the free-fall time, so 
a hot quasi-spherical shell is formed above the magnetosphere with 
temperature determined by hydrostatic equilibrium \cite{DaviesPringle81},  
\cite{Shakura_ea11}. Accretion of matter through such a shell is subsonic,
so no shock is formed above the magnetosphere. The accretion rate $\dot M$
is now determined by the ability of hot plasma to enter magnetosphere.
This 
is the settling accretion regime. 

\section{Settling accretion regime: theory}

Theory of settling accretion regime was elaborated in \cite{Shakura_ea11}. In this regime, 
the accreting matter subsonically 
settles down onto the rotating magnetosphere forming an extended quasi-static shell.
This shell mediates the angular momentum transfer to/from the rotating NS 
magnetosphere by voscous stresses due to large-scale convective motions and turbulence. 
The settling regime of accretion can be realized for moderate accretion rates 
$\dot M< \dot M_*\simeq 4\times 10^{16}$~g/s. At higher accretion rates, a free-fall gap 
above the neutron star magnetosphere appears due to rapid Compton cooling, and accretion 
becomes highly non-stationary. 

\textbf{Mass accretion rate} through the hot shell is determined by 
mean velocity 
of matter entering the magnetosphere, $u(R_A)=f(u)\sqrt{2GM/R_A}$. The dimensionless factor $f(u)$ is determined by the Compton cooling of plasma above magnetosphere and the critical temperature
for Rayleigh-Taylor instability to develop \cite{ElsnerLamb77}, and is found to be
\begin{equation}
f(u)\approx 0.3 \dot M_{16}^{4/11}\mu_{30}^{-1/11}\,,
\end{equation}
where $\dot M_{16}=\dot M/[10^{16} \hbox{g/s}]$ and $\mu_{30}=\mu/[10^{30} \hbox{G}\,\hbox{cm}^3]$
is the NS magnetic moment. The definition of the Alfven radius
in this case is different from the value used for disk accretion: 
\begin{equation}
R_A\approx 10^9[\hbox{cm}]
\left(\frac{\mu_{30}^3}{\dot M_{16}}\right)^{2/11}\,.
\end{equation} 
As the X-ray luminosity increases, Compton cooling occurs faster, and when $f(u)\to \approx 0.5$, the sonic point appears in the accretion flow above the magnetosphere, and the accretion becomes highly non-stationary. Thus the condition $f(u)=0.5$ determines the critical X-ray luminosity for the steady settling accretion: $L_x<4\times 10^{36}\mu_{30}^{1/4}$~erg/s.   

\begin{figure}
\includegraphics[width=0.45\textwidth]{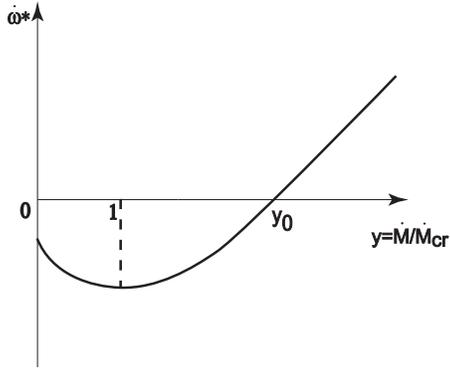}
\caption{A schematic plot of $\dot\omega^*$ as a function of $y$.
Torque reversal occurs at $y_0$. Sign of torque-luminosity correlations
changes at $y=1$.}
\label{f:y}
 \end{figure}

\textbf{Accretion torques} applied to NS in this regime are determined not only by the specific angular momentum of captured matter $j_m$ (as is the case
of the free-fall accretion), but also by the possibility to transfer angular momentum to/from 
the rotating magnetosphere through the shell by large-sclae convective motions. This means that the NS spin-down becomes possible even for prograde $j_m$. The plasma-magnetosphere interaction results in emerging of the toroidal magnetic field 
$B_t=K_1 B_p(\omega_m-\omega^*)t_{inst}$, where $\omega_m$ is the angular frequency
of matter at the Alfven radus, $K_1$ the dimensionless coupling coefficient which is different in 
different sources, $t_{inst}$ is the characteristic time of RT instability. The gas-dynamical treatment of the problem of angular momentum transfer through the shell by viscous turbulence stresses \cite{Shakura_ea11} showed that $\omega_m\approx \Omega_b(R_A/R_B)^n$, where the index 
$n$ depends on the character of turbulence in the shell. For example, in the case of isotropic near-sonic turbulence we obtain $n\simeq 3/2$, i.e. quasi-Keplerian rotation distribution. In the more likely case of strongly anisotropic turbulence (because of strong convection) we find $n\approx 2$, i.e. an iso-angular-momentum distribution.  
The NS spin evolution equation reads:
\begin{equation}
\label{sd_eq2}
I\dot \omega^*=A\dot M^{\frac{3+2n}{11}} - B\dot M^{3/11}\,, 
\end{equation} 
where $I$ is the NS moment of inertia. The (independent of $\dot M$) coefficients $A$ and $B$ for the case $n=2$ are (in CGS units):
\begin{equation}
A\approx 5.3\times 10^{31} K_1 \mu_{30}^{\frac{1}{11}}\left(\frac{v}{1000
\hbox{km/s}}\right)^{-4}\left(\frac{P_b}{10\hbox{d}}\right)^{-1}, \qquad
B\approx 5.4\times 10^{32}K_1\mu_{30}^{\frac{13}{11}}\left(\frac{P^*}{100\hbox{s}}\right)^{-1}
\end{equation}
The function $\dot\omega^*(\dot M$) reaches minimum at 
some $\dot M_{cr}$. By differentiating Eq. (\ref{sd_eq2}) with respect to $\dot M$ and equating to zero, we find
$
\dot M_{cr}=\left[\frac{B}{A}\frac{3}{(3+2n)}\right]^{\frac{11}{2n}}
$.
At $\dot M=\dot M_{cr}$ the value of $\dot\omega^*$ reaches an absolute minimum (see Fig. \ref{f:y}). In terms of the dimensionless mass accretion rate
$
y\equiv \frac{\dot M}{\dot M_{cr}}
$
Eq. (\ref{sd_eq2}) can be written in the form
\begin{equation}
\label{sdy}
I\dot \omega^*=A\dot M_{cr}^{\frac{3+2n}{11}}y^{\frac{3+2n}{11}}
\left(1- \left(\frac{y_0}{y}\right)^\frac{2n}{11}\right)\,,
\end{equation}
where the frequency derivative vanishes at $y=y_0$:
$
y_0=\left(\frac{3+2n}{3}\right)^{\frac{11}{2n}}.
$
The qualitative behaviour of $\dot\omega^*$ as a function of $y$ is shown in Fig. 
\ref{f:y}. So in the settling accretion regime transitions from spin-up ($y>y_0$)
to spin-down ($y<y_0$) occurs  with changing $L_x$.
By varying Eq. (\ref{sdy}) with respect to $y$ we find:
\begin{equation}
\label{variations}
I(\delta\dot\omega^*)=I\frac{\partial \dot \omega^*}{\partial y}(\delta y)=
\frac{3+2n}{11}A\dot M_{cr}^\frac{3+2n}{11}y^{-\frac{8-2n}{11}}
\left(1- \frac{1}{y^{\frac{2n-1}{11}}}\right)(\delta y)\,.
\end{equation}
We see that depending on whether $y>1$ or $y<1$, 
\textit{correlated changes} of $\delta \dot\omega^*$
with X-ray flux should have different signs. Indeed, for GX 1+4 in
\cite{GG11a}, \cite{GG11b} 
a positive correlation
of the observed $\delta P$ with $\delta \dot M$ was found using \textit{Fermi} data
(see also \cite{Chakrabarty_ea97} for BATSE data). This means that 
there is a negative correlation between $\delta\omega^*$ and $\delta\dot M$, suggesting $y<1$ in this source.

\section{Settling accretion regime: observations}

First, the settling accretion regime explains long-term spin-down states as
observed, for 
example, in GX 1+4 \cite{GG11b}.
Another immediate application of Eq. ({\ref{sd_eq2}) to observations 
is the estimation of the NS magnetic field in XPSRs 
with equilibrium spin period (i.e. where $\dot \omega^*=0$ on average):
\begin{equation}
\label{mu_eq}
\mu_{30}^{(eq)}\approx 0.3
\left(\frac{P_*/100s}{P_b/10d}\right)^\frac{11}{12}\dot M_{16}^\frac{1}{3}
\left(\frac{v}{1000 \hbox{km/s}}\right)^{-\frac{11}{3}}
\end{equation}
(here we assumed $n=2$ and maximum anisotropic turbulence). In wind-fed pulsars Vela X-1 and GX 301-2, 
the NS magnetic field estimated in this way was found to be close to the value inferred 
from the cyclotron line measurements (see \cite{Shakura_ea11} for more detail). 

This equation also 
implies that the equilibrium period of XPSRs in the settling accretion regime is 
\begin{equation}
\label{P_eq}
P_{eq}\approx 1000 [\hbox{s}]\mu_{30}^{12/11}\dot M_{16}^{-4/11}\left(\frac{P_b}{10\hbox{d}}\right)
\left(\frac{v}{1000 \hbox{km/s}}\right)^4.
\end{equation}
There are known several very slowly rotating XPSRs, including some 
in HMXB (4U 2206+54 with $P^*=5550$ s), 2S 0114+65 with $P^*=9600$~s) and some in 
LMXB (e.g. 3A 1954+319, $P^*=19400$~s \cite{Marcu_ea11}). 
Assuming disk accretion in such pulsars would 
require incredibly high NS magnetic fields.  But in 4U 2206+54 the NS field is measured
to be normal \cite{Wang09} , $B\simeq 3\times 10^{12}$~G (in 2S 0114+65 possibly too, 
with $B\sim 2.5\times 10^{12}$~G \cite{BonningFalanga05}). 
X-ray luminosity in these pulsars is low, 
$\sim 10^{36}$~erg/s, which suggests the settling accretion regime in these pulsars. 

Another possible implication of the settling accretion theory can be for non-stationary 
phenomena in XPSRs. For example,  
a dynamical instability of the shell on 
the time scale of the order of the free-fall time from the magnetosphere 
can appear due to increased Compton cooling
and hence increased mass accretion rate in the shell. This  
may even result in a complete collapse of the shell
resulting in an X-ray outburst with duration similar to 
the free-fall time scale of the entire shell
($\sim 1000$~s). Such a transient behaviour 
is observed in supergiant fast X-ray transients (SFXTs) (see e.g. \cite{Ducci_ea10}). 
Slow X-ray pulsations are found in some of them (e.g. IGRJ16418-4532, $P^*\approx 1212$~s
\cite{Sidoli_ea11}). In this source, the regular pulsations distinctly seen at a small 
X-ray luminosity $L_x\sim 10^{34}$~erg/s were found to disappear when the average 
X-ray flux increased by more than an order of magnitude, and quasi-regular  
flares were observed on a time scale of 200-250 s. In such a low-luminosity source the 
magnetospheric radius should be very large, $R_A\simeq 5\times 10^{10}$~cm (assuming
the standard value for the NS magnetic field $\mu_{30}=1$). So the observed flaring 
behavior can be the manifestation of a 
Rayleigh-Taylor instability from the magnetospheric radius which takes place
on the dynamical time scale $\sim R_A^{3/2}/\sqrt{GM}$. Note that at small 
X-ray luminosities 
the radiation cooling (and not the Compton cooling) 
controls plasma entering the magnetosphere, so the critical luminosity 
for unstable accretion becomes
lower than $4\times 10^{36}$~erg/s found for the Compton cooling case.

\section{Conclusion}
        
At X-ray luminosities $<4\times 10^{36}$~erg/s wind-fed
X-ray pulsars can be at the stage of subsonic
settling accretion. In this regime, accretion rate
onto NS is determined by the ability of plasma
to enter magnetosphere via Rayliegh-Taylor instability. 
A gas-dynamic theory of settling accretion
regime is constructed taking into account
anisotropic turbulence \cite{Shakura_ea11}. The angular momentum can be
transferred through the quasi-static shell via
large-scale convective motions initiating turbulence cascade. 
The angular velocity distribution in the shell is found
depending on the turbulent viscosity prescription.
Comparison with observations of long-period X-ray
wind-fed pulsars GX 301-2 ad Vela X-1 shows that  an almost
iso-angular-momentum distribution is most likely realized in their shells. 
The theory explains long-term spin-down in wind-
fed accreting pulsars and properties of short-term
torque-luminosity correlations. Long-period low-luminosity X-ray pulsars are most
likely experiencing settling accretion too. 
Spectral and
timing measurements can be used to further test this accretion regime.

\end{document}